\documentclass[trackchanges]{aastex631}
\shorttitle{}
\shortauthors{Pozo Nu\~nez et al.}
\graphicspath{{./}{figures/}}

\begin{document}

\title{Reevaluating LSST's Capability for Time Delay Measurements in Quasar Accretion Discs}

\author[0000-0002-6716-4179]{F. Pozo Nu\~nez}
\affiliation{Astroinformatics, Heidelberg Institute for Theoretical Studies, Schloss-Wolfsbrunnenweg 35, 69118 Heidelberg, Germany}

\author[0000-0001-5848-4333]{B. Czerny}
\affiliation{Center for Theoretical Physics, Polish Academy of Sciences, Al. Lotnik\'ow 32/46, 02-668 Warsaw, Poland}

\author[0000-0002-5854-7426]{S. Panda}\thanks{CNPq fellow}
\affiliation{Laborat\'orio Nacional de Astrof\'isica - MCTI, R. dos Estados Unidos, 154 - Na\c{c}\~oes, Itajub\'a - MG, 37504-364, Brazil}

\author[0000-0001-5139-1978]{A. Kovacevic}
\affiliation{University of Belgrade-Faculty of Mathematics, Department of Astronomy, Studentski trg 16, Belgrade, Serbia}

\author[0000-0002-0167-2453]{W. Brandt}
\affiliation{Department of Astronomy and Astrophysics, 525 Davey Lab, The Pennsylvania State University, University Park, PA 16802, USA}

\author[0000-0003-1728-0304]{K. Horne}
\affiliation{School of Physics and Astronomy, University of St Andrews, North Haugh, St Andrews, KY16 9SS, Scotland, UK}

\collaboration{10}{on behalf of the LSST AGN Science Collaboration}



\begin{abstract}

The Legacy Survey of Space and Time (LSST) at the Vera C. Rubin Observatory is poised to observe thousands of quasars using the Deep Drilling Fields (DDF) across six broadband filters over a decade. Understanding quasar accretion disc (AD) time delays is pivotal for probing the physics of these distant objects. \cite{2023MNRAS.522.2002P} has recently demonstrated the feasibility of recovering AD time delays with accuracies ranging from 5\% to 20\%, depending on the quasar's redshift and time sampling intervals. Here we reassess the potential for measuring AD time delays under the current DDF observing cadence, which is placeholder until a final cadence is decided.

We find that contrary to prior expectations, achieving reliable AD time delay measurements for quasars is significantly more challenging, if not unfeasible, due to the limitations imposed by the current observational strategies.

\end{abstract}




\section{Introduction} \label{sec:intro}

Reverberation mapping sheds light on the spatial extent and temperature stratification of the AD, offering crucial insights for understanding black hole properties and its application in cosmology (\citealt{1999MNRAS.302L..24C}; \citealt{2017PASP..129i4101P}; \citealt{2021iSci...24j2557C}; \citealt{2023Ap&SS.368....8C}). Given this, the upcoming LSST will be vital due to its ability to deliver an extensive collection of quasar light curves across six broadband filters (Sloan-$u, g, r, i, z, y$) over ten years. Consequently, the community must evaluate its performance and update predictions before the start of the survey.

\section{Time delay recovery}

We adapted the modeling approach outlined in \cite{2023MNRAS.522.2002P} (hereafter PN23) for light curves, conforming to the DDF observing cadence. 
As a specific example, we used the baseline DDF cadence\footnote{This refers to the {ddf\_v1.7\_nodither\_10yrs} for the ELAIS-S1 field centered at RA=9.45$^{\circ{}}$, Dec = -44.025$^{\circ{}}$. We refer the readers to \citet{2023A&A...675A.163C} where the performance of multiple LSST cadence strategies are evaluated.} from \citet{2023A&A...675A.163C}.
Four test cases ("source type") were considered, differentiated by redshift, black hole mass, and accretion rate, to develop the AD transfer functions (see Table 1 in PN23). For each test case, we generated 2000 random light curves spanning 3700 days, each with a signal-to-noise ratio of 100 in each filter. 
Time delays relative to the $u$-band (the driver light curve) were calculated using ICCF (\citealt{1987ApJS...65....1G}), and Von-Neumann (\citealt{2017ApJ...844..146C}) methods. 
However, for brevity, we present only the ICCF results, due to its widespread use and comparable performance. 

\begin{figure*}
\begin{tabular}{cc}
  \includegraphics[width=70mm]{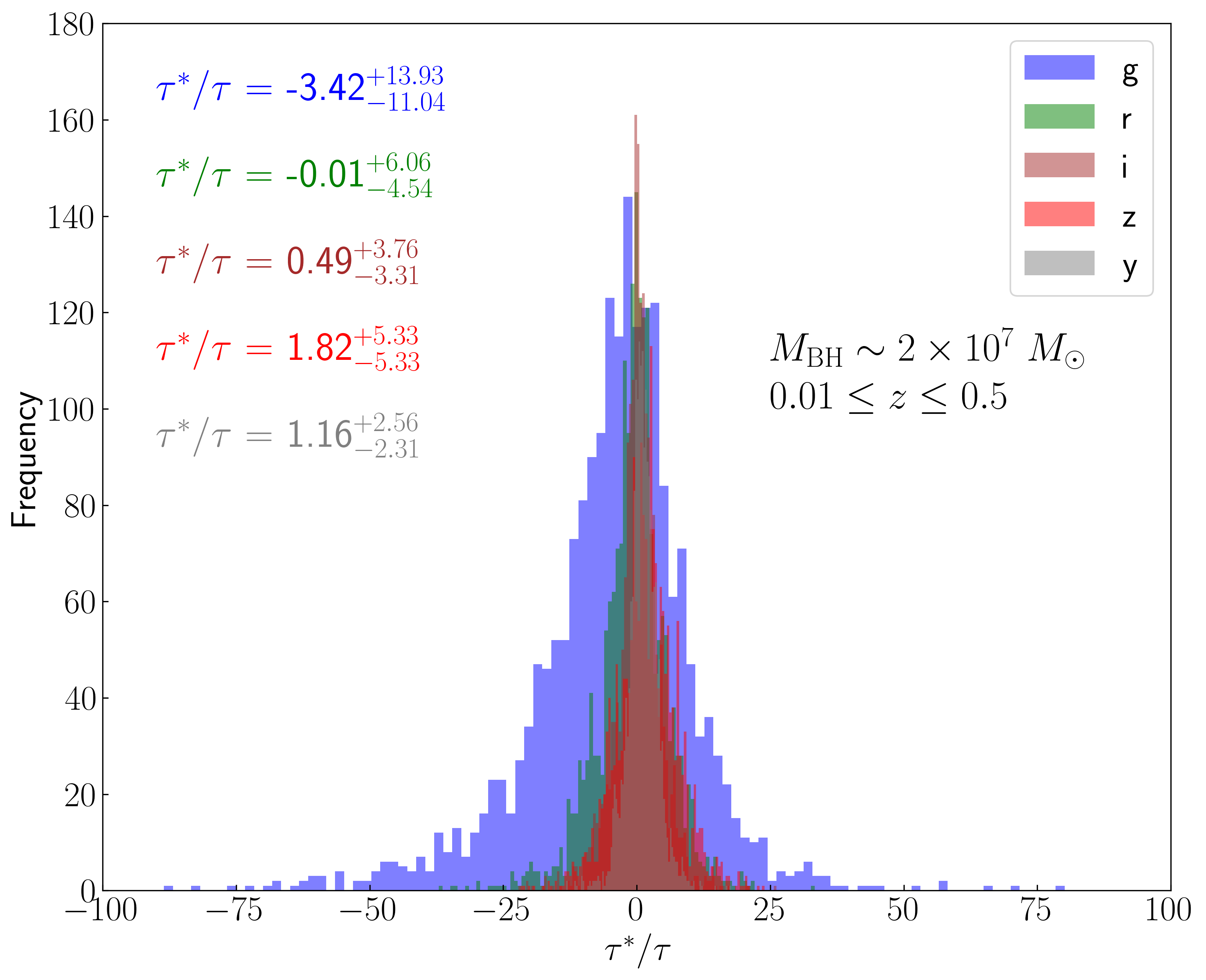} &   \includegraphics[width=70mm]{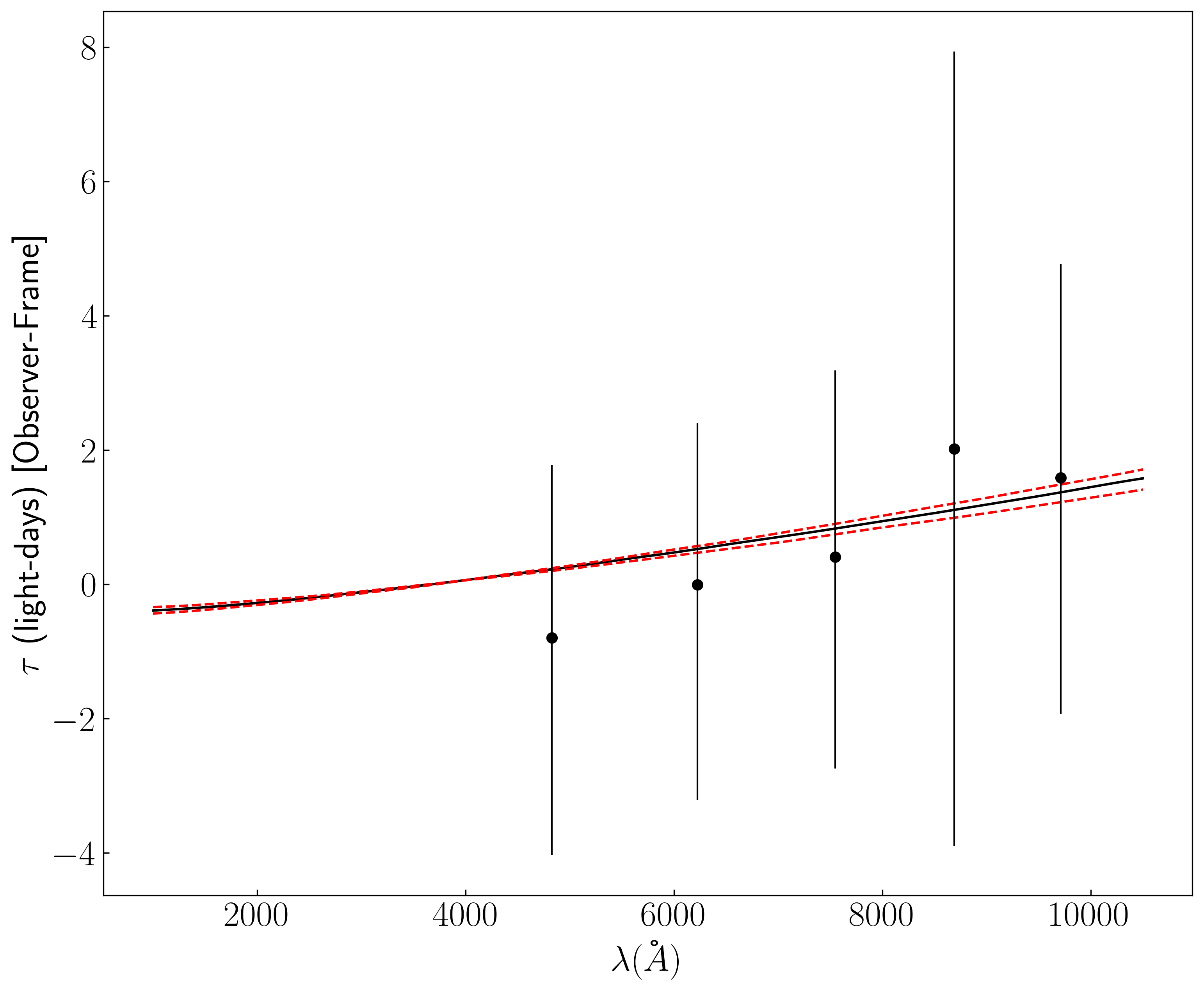}  \\ 
  \includegraphics[width=70mm]{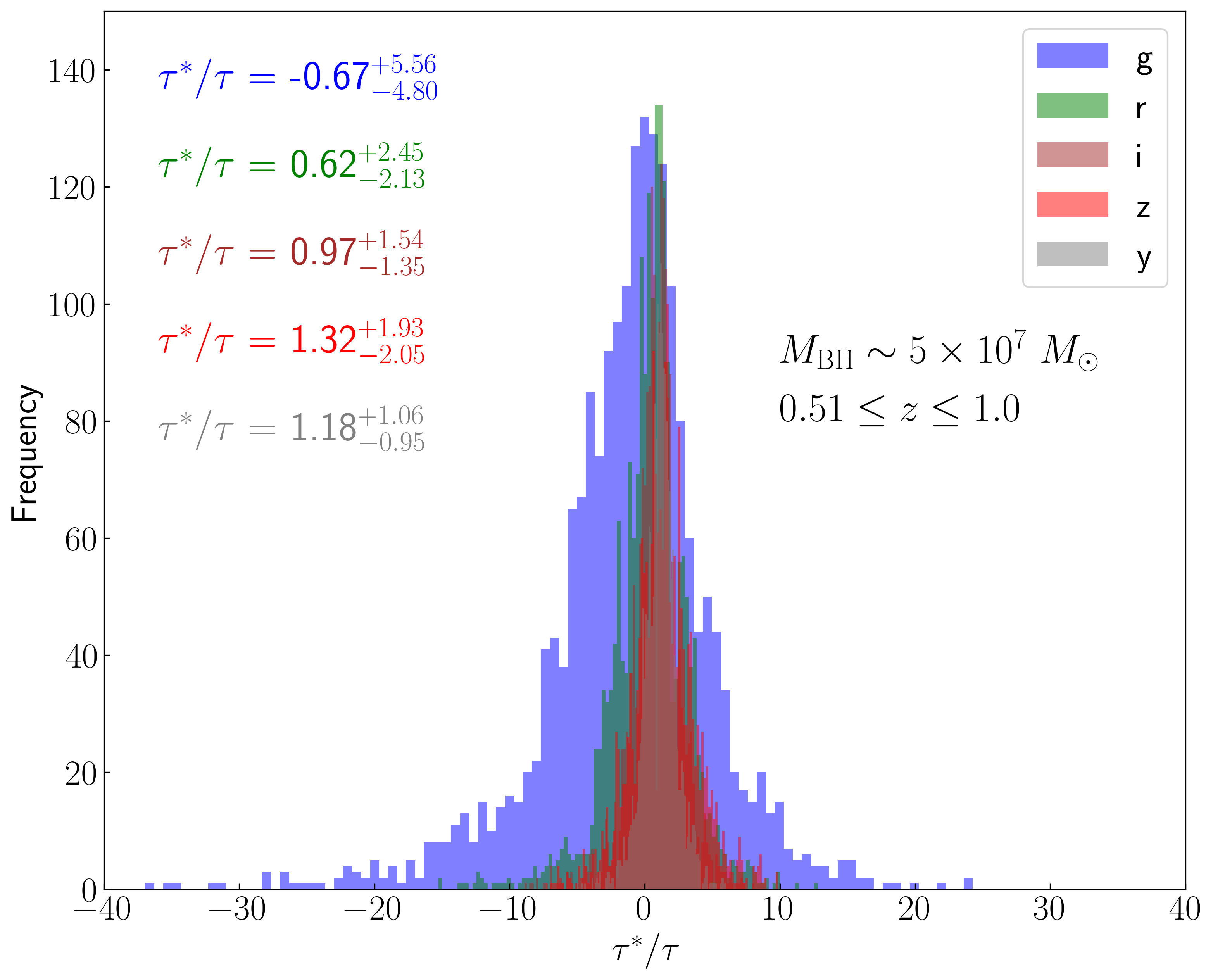} &   \includegraphics[width=70mm]{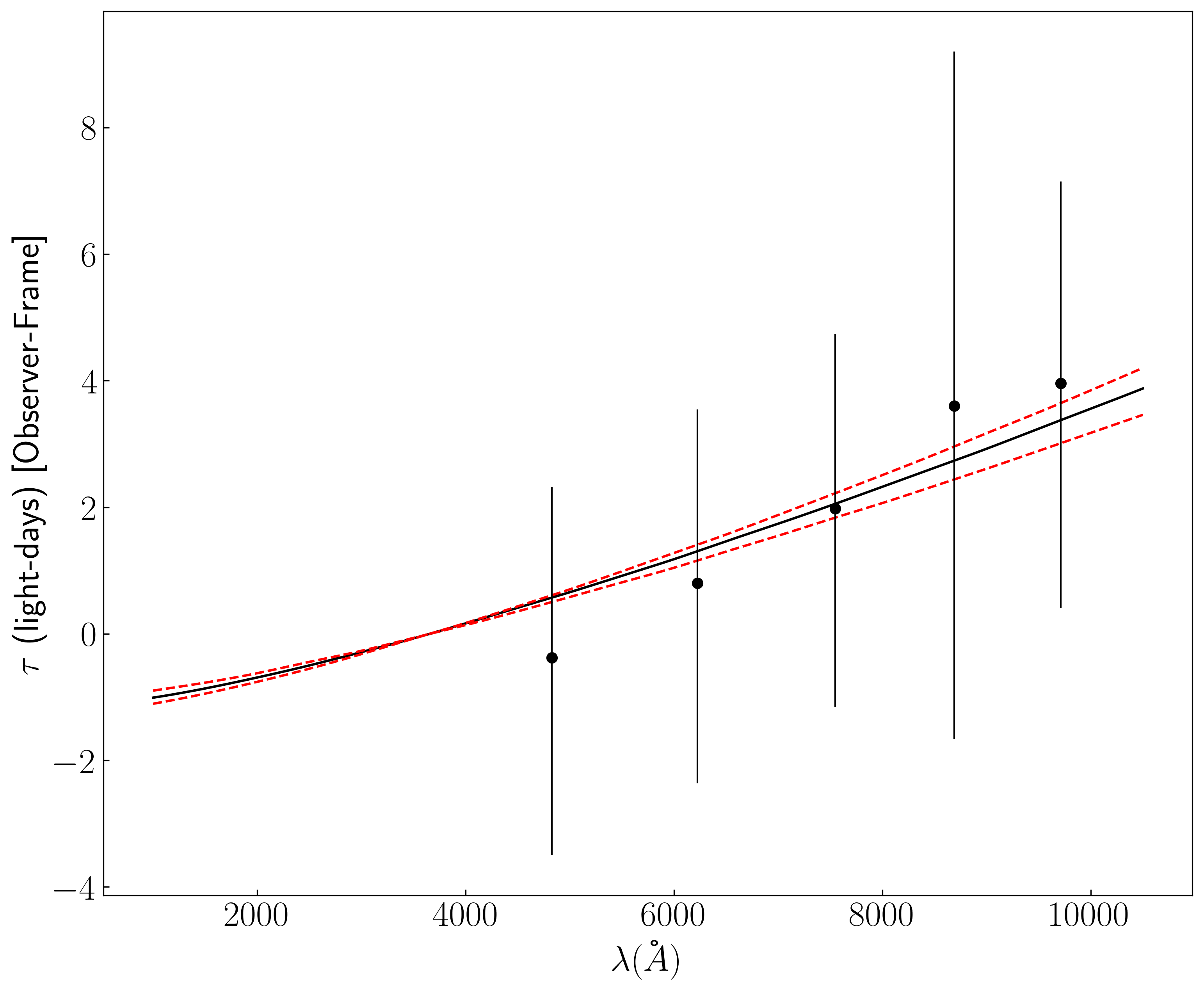}  \\
  \includegraphics[width=70mm]{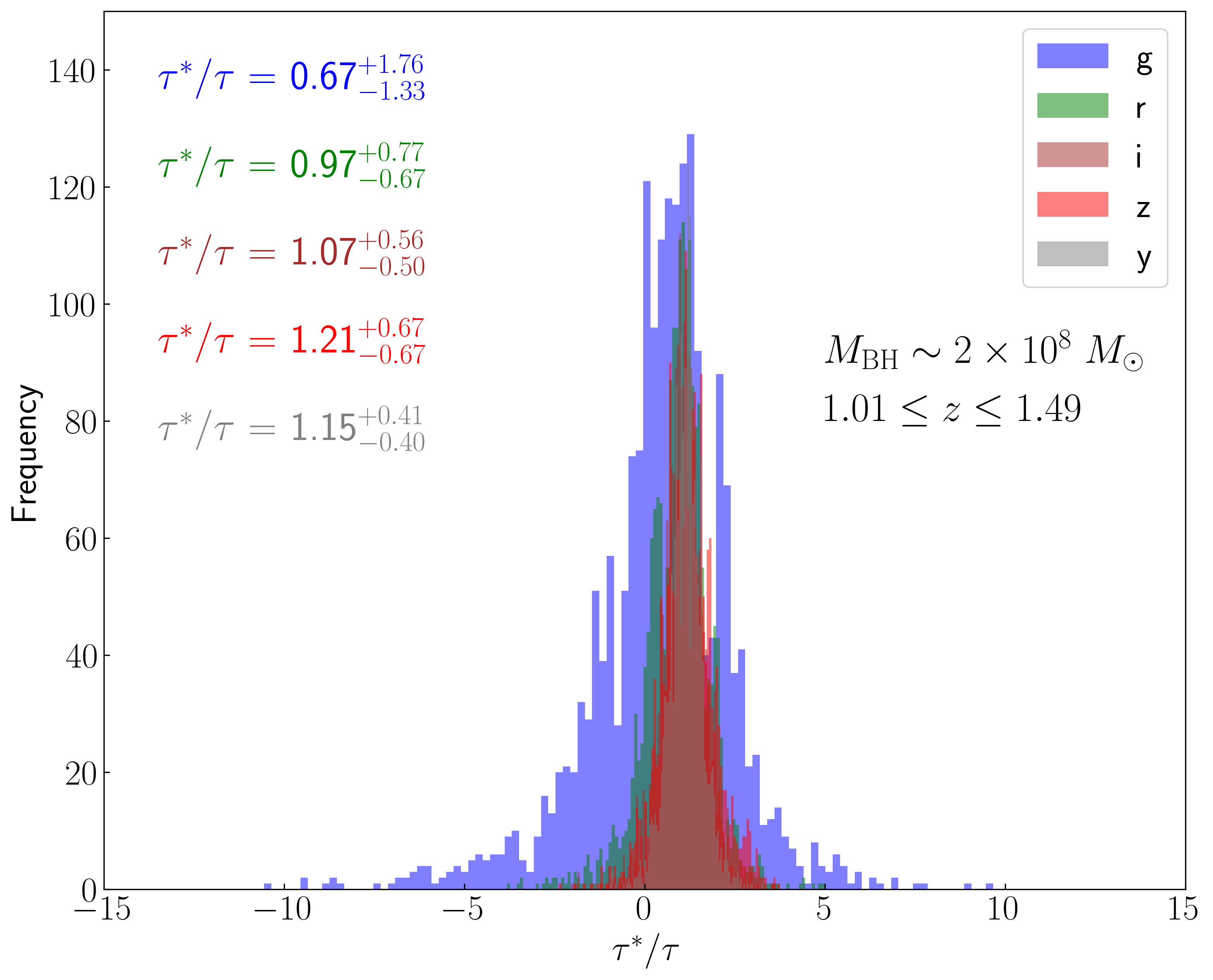} &  \includegraphics[width=70mm]{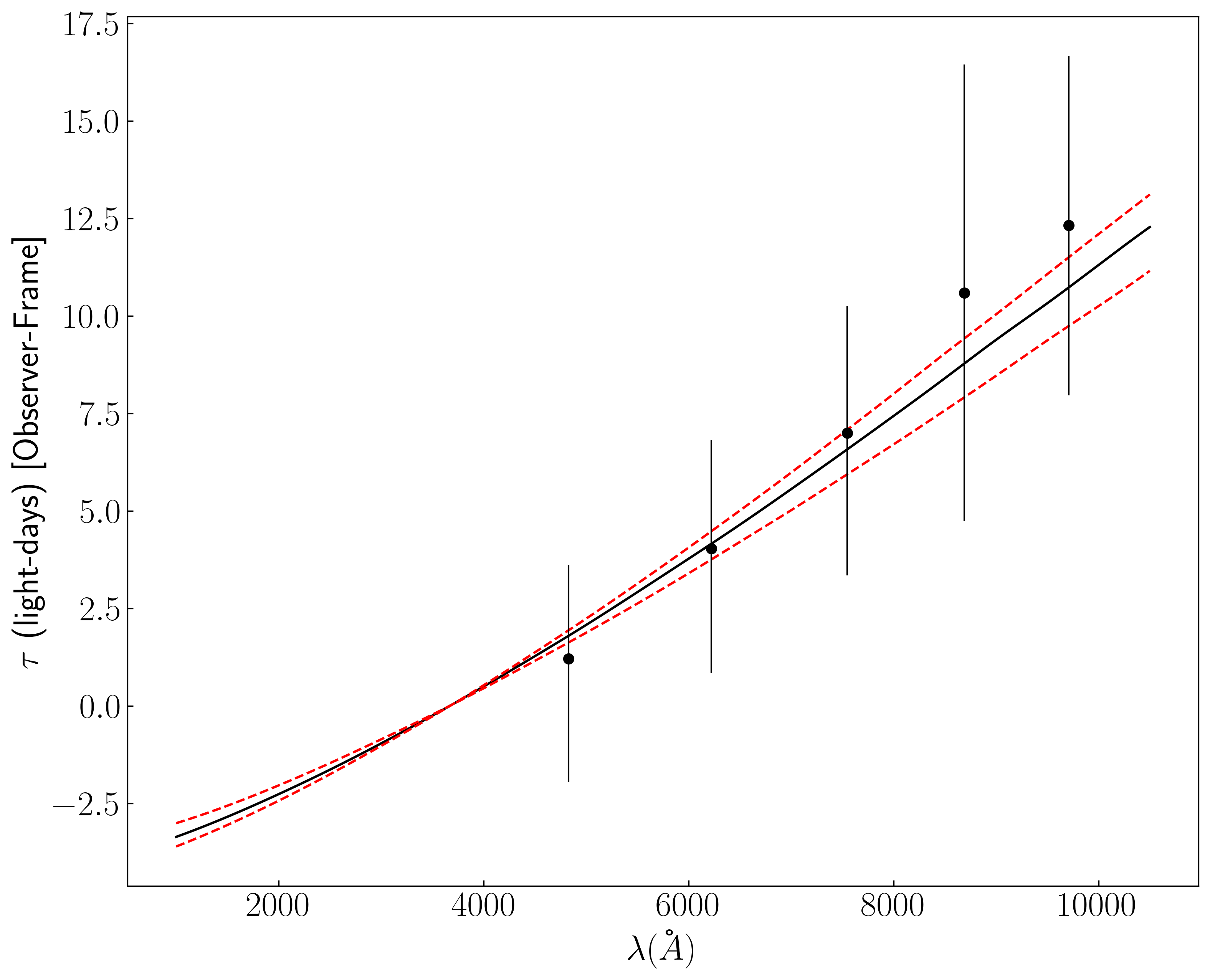}  \\
  \includegraphics[width=70mm]{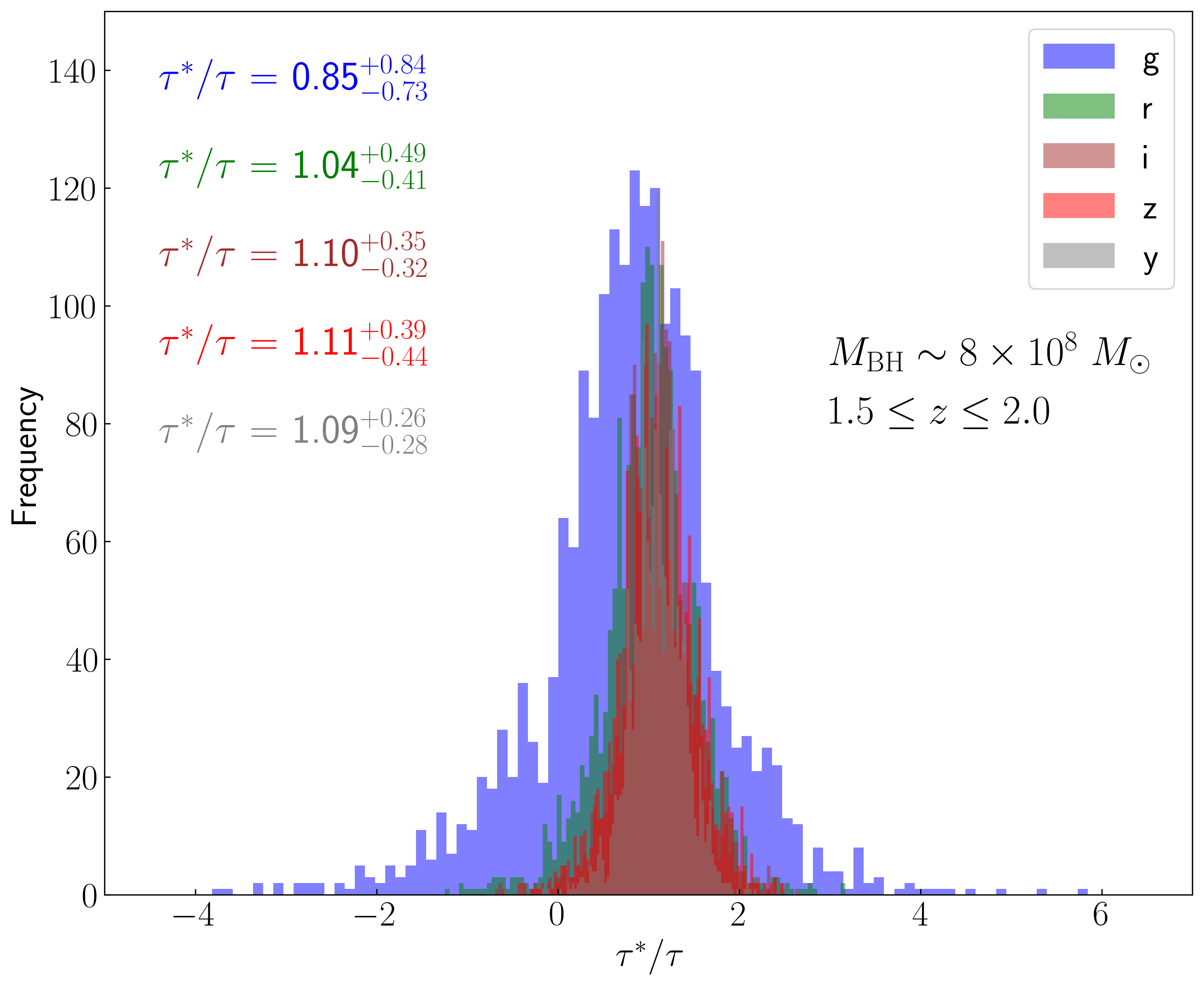} &  \includegraphics[width=70mm]{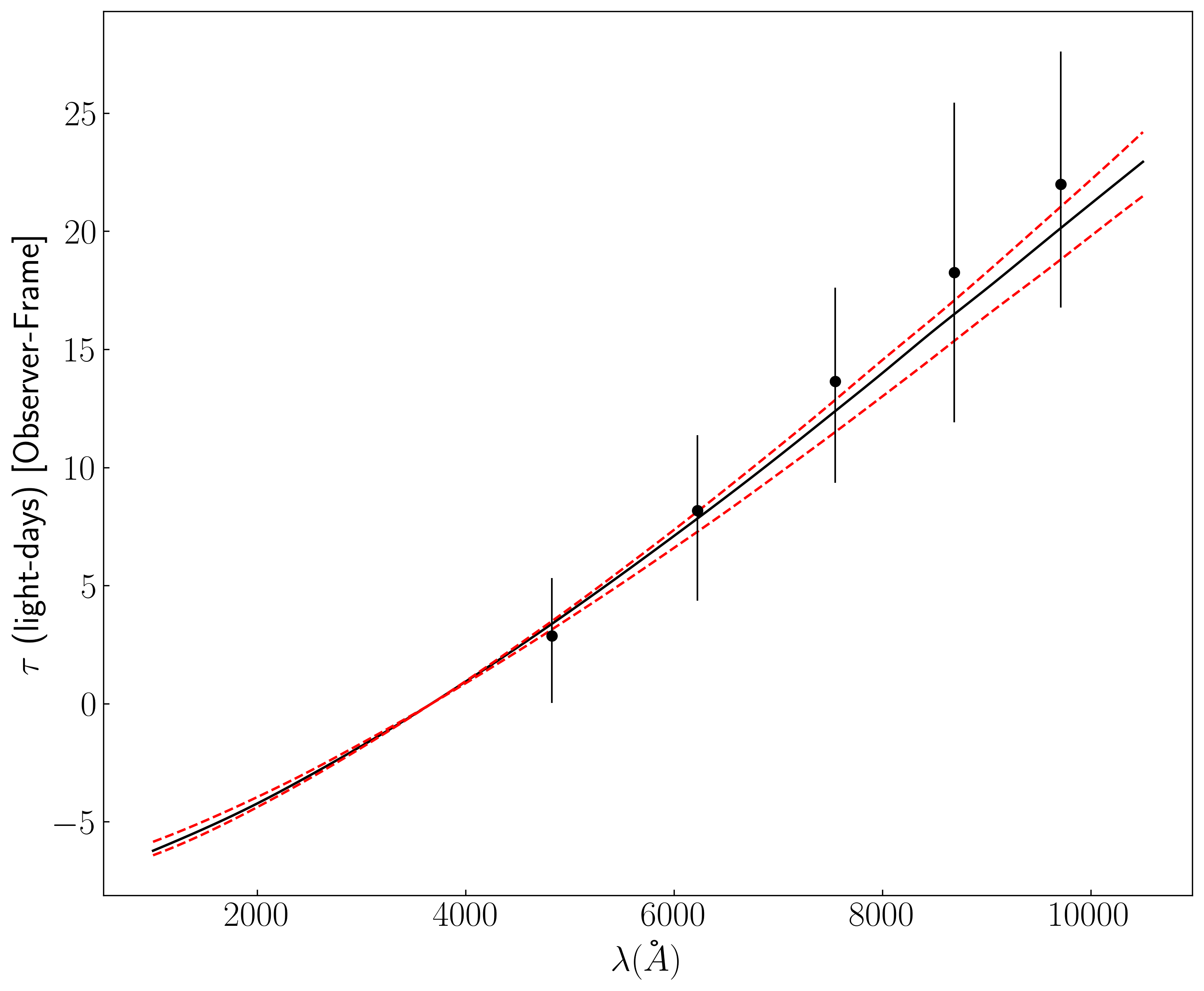} \\
\end{tabular}
\caption{\textit{Left:} Recovered distributions of delays ($\tau^{*}$) for different redshift and black hole masses. \textit{Right:} time delay spectrum $\tau_{c}(\lambda)\propto \lambda^{4/3}$ (black line) as predicted from the response functions. The black circles represent the mean delays and the error bars the standard deviation expected for a single source measurement. The dotted red lines show the delay spectrum obtained for a black hole mass with $30$\% uncertainty.}
\label{fig:fig1}
\end{figure*}

Figure \ref{fig:fig1} shows the recovered distributions of delays ($\tau^{*}$) highlighting the median and the central 68\% confidence (1$\sigma$) interval. It is evident that for lower redshifts and smaller black hole masses ($\leq \sim5\times10^{7}$ M${\odot}$; first two plots from the top), recovering the delay is problematic, and uncertainties are substantial. This is primarily due to the light curve's undersampled nature, deviating significantly from the Nyquist theorem. This effect is more pronounced at wavelengths closer to the driver light curve, as the delay difference is smaller, sometimes even indicating negative delays (e.g., $g$ and $r$ bands). The situation improves slightly at higher redshifts and for larger black hole masses ($\sim10^{8} - 10^{9}$ M${\odot}$; bottom plot), where the time delay is more significant, and sampling more closely aligns with the Nyquist theorem. In this scenario, the best-case recovery is for $i$, $z$, and $y$ bands, with uncertainties around 30\%. For $g$ and $r$ bands, uncertainties are approximately 90\% and 40\%, respectively.

\section{Future}

As outlined in PN23, time delays for quasars at redshifts \(1.5 < z < 2.0\) can be recovered with an accuracy between 5\% to 10\% for a time sampling interval of 2 to 5 days. This analysis applies to black hole masses in the range of approximately \(10^8\) to \(10^9\) M\(_{\odot}\).
However, recovering the time delay for smaller black hole masses, specifically those below \(5 \times 10^7\) M\(_{\odot}\), is not feasible with the current baseline DDF cadence. To extend this capability to lower mass ranges, a denser time sampling interval of 1 to 2 days is required for all bands and without significant gaps (light curve duration of at least twice the time delay to be detected), which could improve the precision of recovered delays to about 10 to 20\%. Consequently, for effective quasar AD time delay science, we recommend that the DDF should focus on shorter monitoring periods, such as a single season of 6 months, with a time sampling interval of 1 day, as similarly proposed in previous experiments (\citealt{2020ApJS..246...16Y}; PN23). This aligns with the recommendations presented by the AGN Science Collaboration's \textit{accordion} cadence note\footnote{\url{https://docushare.lsst.org/docushare/dsweb/Get/Document-37655/agn-ddf-cadence-note01.pdf}}. This \textit{accordion} cadence is proposed to be dense and uniform, but only over the period of 2.5 months in a given year, in order to preserve the total-depth measurement requirements, while PN23 claims very satisfactory results assuming six months for the duration of the monitoring. As the question of cadence has not yet been resolved, it will be important to examine later in detail the proposition from new developments in cadence optimisation.

\begin{acknowledgments}
Author F. Pozo Nu\~nez gratefully acknowledges the generous and invaluable support of the Klaus Tschira Foundation. This project has received funding from the European Research Council (ERC) under the European Union's Horizon 2020 research and innovation program (grant agreement No 951549). S. Panda acknowledges the financial support of the Conselho Nacional de Desenvolvimento Científico e Tecnológico (CNPq) Fellowships 300936/2023-0 and 301628/2024-6.
\end{acknowledgments}

%








\bibliography{fplsst}{}
\bibliographystyle{aasjournal}



\end{document}